\documentclass{article} % For LaTeX2e
\usepackage{iclr2026_conference,times}

\usepackage{amsmath,amsfonts,bm}

\def\eqref#1{equation~\ref{#1}}
\def\1{\bm{1}}

\DeclareMathAlphabet{\mathsfit}{\encodingdefault}{\sfdefault}{m}{sl}
\SetMathAlphabet{\mathsfit}{bold}{\encodingdefault}{\sfdefault}{bx}{n}

\usepackage{hyperref}
\usepackage{url}
\usepackage{booktabs}

\usepackage{amsmath}
\usepackage{amsfonts}
\usepackage{amssymb}

\usepackage{algorithm}
\usepackage{algorithmic}

\usepackage{xcolor}

\usepackage{graphicx}
\usepackage{float}
\usepackage{multirow}
\usepackage{subfig}

\usepackage{array}
\usepackage{etoolbox}

\robustify{\subref}

\title{Adaptive KV-Cache Compression without Manually Setting Budget}

\author{Chenxia Tang, Jianchun Liu, Hongli Xu, Liusheng Huang}

\iclrfinalcopy % Uncomment for camera-ready version, but NOT for submission.
\begin{document}

\maketitle

\begin{abstract}
Large language models (LLMs) inference relies heavily on KV-caches to accelerate autoregressive decoding, but the resulting memory footprint grows rapidly with sequence length, posing significant efficiency challenges.
Current KV-cache compression methods suffer from a Procrustes' bed problem: they force diverse workloads into fixed compression ratios, leading to suboptimal resource allocation and inference performance. 
To this end, we present GVote, an adaptive KV-cache compression scheme that eliminates manual budget specification while achieving superior accuracy-efficiency trade-offs. 
GVote operates on the principle that the important keys are the aggregation of keys required by future queries. 
The method predicts future query attention demands by Monte-Carlo style sampling potential queries and aggregating selected keys to determine the optimal cache budget without manual specification.
Experimental evaluation demonstrates GVote's effectiveness across multiple benchmarks, including GSM8K, RULER and Longbench. 
Compared to baselines, GVote exhibits 2$\times$ memory reduction while the accuracy maintains higher or comparable.
\end{abstract}

\section{Introducion}
Large language models (LLMs) rely on key–value (KV) caches to store intermediate attention results during autoregressive decoding, enabling efficient reuse of previously computed representations.
However, the deployment of LLMs faces a critical bottleneck: the quadratic growth of KV-cache memory with sequence length~\citep{attention}. 
In practical deployments, the KV-cache often dominates GPU memory usage, limiting batch size, inflating inference latency, and restricting model accessibility in resource-constrained environments.
Consequently, efficient KV-cache compression has become essential for enabling long-context reasoning while maintaining the accuracy and efficiency of LLM inference.~\citep{longformer,bigbird}.

Recent advances such as SnapKV and AdaKV~\citep{snapkv,adakv} demonstrate that exploiting the sparsity of attention scores can yield substantial KV-cache compression.
However, these approaches follow a rigid \textit{fixed-budget} paradigm, where practitioners must pre-allocate a static memory quota (e.g., retaining 20\% of the cache) without foreknowledge of the contextual demands of incoming requests.
Such a \textit{one-size-fits-all} design inevitably forces heterogeneous workloads into the same compression ratio, leading to a Procrustes’ bed problem that creates an intractable trade-off between memory efficiency and model accuracy.

\begin{figure}[t]
\centering
\includegraphics[width=\textwidth]{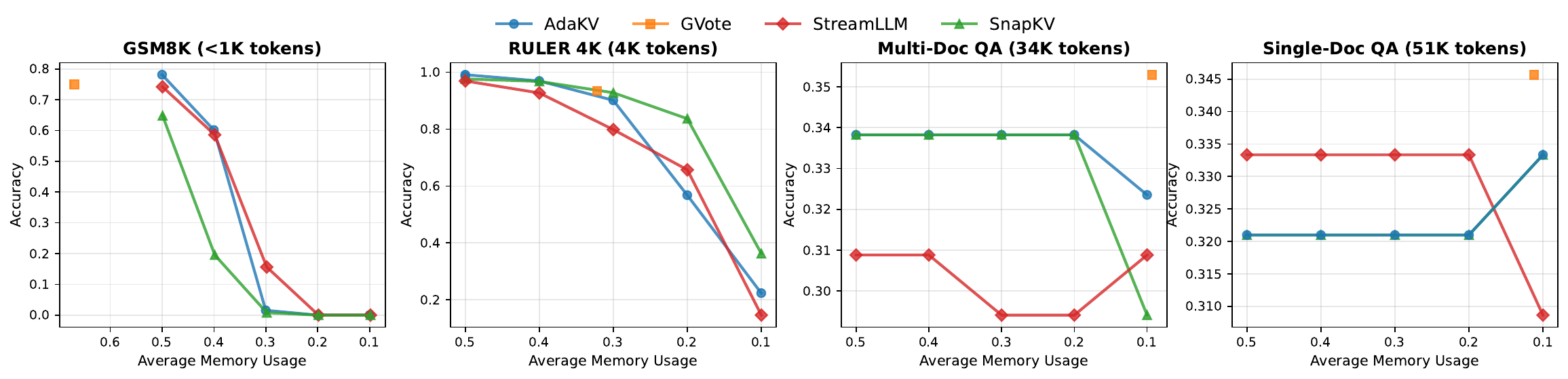}
\caption{Accuracy-usage analysis across diverse benchmarks using the Llama-3.1-8B-Instruct model. The four panels display results for GSM8K (\textless 1K tokens), RULER-4K (4K tokens),  Multi-Doc QA (34K tokens) and Single-Doc QA (51K tokens), representing tasks with varying token lengths. The graphs show Accuracy on the y-axis against Average Memory Usage on the x-axis. The performance of StreamLLM, SnapKV, AdaKV, and GVote (ours) is compared. GVote consistently finds a favorable trade-off between accuracy and memory usage, achieving good accuracy while maintaining low memory consumption, thereby eliminating the need for manual budget settings, as shown in the plots.}
\label{fig:accuracy_usage_analysis}
\end{figure}

Consider a production LLM inference engine serving diverse requests, ranging from mathematical reasoning (e.g., GSM8K~\citep{gsm8k}) to long-document analysis (e.g., RULER-4K~\citep{ruler}) and various QA tasks.
When the cache budget is set too low (e.g., 20\%), memory efficiency is improved but reasoning-intensive tasks suffer catastrophic degradation, with accuracy dropping to nearly zero.
Conversely, allocating a higher budget (e.g., 50\%) preserves performance on complex tasks but results in substantial memory waste for simpler workloads, where accuracy remains stable even under lower budgets, as illustrated in Figure~\ref{fig:accuracy_usage_analysis}.

This fundamental conflict—where any fixed compression ratio risks either severe performance degradation or significant memory inefficiency—makes fixed-budget approaches inherently unsuitable for dynamic workloads.
Beyond this core limitation, such methods also impose expensive dataset-specific hyperparameter tuning~\citep{automl}, since the optimal cache budget varies across tasks and domains.
They further exhibit brittleness under distribution shifts, where a budget tuned for one workload may fail catastrophically on another.
Finally, to guard against worst-case scenarios, practitioners are forced to adopt conservative over-provisioning strategies, which squander valuable memory resources and exacerbate deployment costs.

To this end, we introduce GVote, an adaptive per-request KV-cache compression scheme that automatically computes optimal cache budgets without manual specification. 
GVote's approach is based on our observation that hidden states exhibit a Gaussian distribution. 
Building on this finding, we sample a small number of states, S, from this distribution, compute their corresponding queries, and then form the final KV-cache by taking the union of all required keys.

GVote delivers substantial empirical improvements over fixed-budget methods across diverse benchmarks. 
As illustrated in Figure~\ref{fig:accuracy_usage_analysis}, different datasets exhibit dramatically different optimal compression ratios, making fixed-budget approaches fundamentally inadequate. 
The four panels demonstrate that conservative budgets (50\%+) maintain accuracy but waste memory on simple tasks, while aggressive budgets (20\%-) collapse accuracy on complex tasks. In contrast, GVote automatically computes the optimal accuracy-efficiency operating point for each request. 
For instance, on the Multi-Doc QA benchmark, GVote achieves approximately 0.35 accuracy with only 10\% average memory usage, while other methods require at least double the memory for lower accuracy. 
This highlights how no single fixed budget can efficiently serve diverse workloads, whereas GVote's adaptive computation eliminates this fundamental limitation.

Our contributions are threefold. 
First, we identify and formalize the fundamental limitation of fixed-budget KV-cache compression in serving diverse workloads. 
Second, we propose GVote, a novel adaptive compression algorithm that automatically computes optimal cache budgets through query sampling and voting mechanisms.
Third, we demonstrate substantial empirical improvements across multiple benchmarks, showing that adaptive computation significantly outperforms fixed-budget approaches while eliminating manual tuning requirements~\citep{automl}.

\section{Background}

\subsection{Notation and Problem Formulation}

We consider modern autoregressive LLMs comprising $L$ transformer layers, where each layer contains Grouped-Query Attention (GQA)~\citep{GQA} or Multi-Head Attention (MHA)~\citep{attention} with $H_{\text{kv}}$ key-value heads. To accelerate inference, these models employ KV-cache~\citep{kvcache} to store intermediate representations.

During attention computation at step $t$, the model first calculates attention weights:
\begin{equation}
\mathbf{A}_t = \text{softmax}\left(\frac{\mathbf{Q}_t \mathbf{K}_{1:t-1}^T}{\sqrt{d_k}}\right) \in \mathbb{R}^{1 \times (t-1)}
\end{equation}
then performs a weighted aggregation of values:
\begin{equation}
\text{Attention}(\mathbf{Q}_t, \mathbf{K}_{1:t-1}, \mathbf{V}_{1:t-1}) = \mathbf{A}_t \mathbf{V}_{1:t-1}
\end{equation}

Crucially, recent work has revealed that attention weights $\mathbf{A}_t$ exhibit extreme sparsity, with only a small fraction of tokens receiving significant attention~\citep{h2o,snapkv}. This sparsity motivates KV-cache compression: we can selectively retain important key-value pairs without substantially degrading performance.

\textbf{The Top-Down Framework.} Existing compression methods can be formulated as a three-step top-down process: first \emph{scoring}, then \emph{allocation}, and finally \emph{selection}. 
\paragraph{Step 1: Scoring.} 
Each token in the cache is assigned an importance score based on its interaction with queries. 
Formally, a scoring function $\text{Score}(\cdot)$ computes token-level importance:  
\begin{align}
\mathbf{S} &= \text{Score}(\mathbf{K}, \mathbf{V}, \mathbf{Q}) \in \mathbb{R}^{L \times H \times S}, 
\end{align}
where $L$ denotes the number of layers, $H$ the number of heads, and $S$ the sequence length.  

\paragraph{Step 2: Allocation.} 
A global budget $B$ is divided across different attention heads. 
The allocation function $\text{Alloc}(\cdot)$ distributes the budget into per-head quotas $b^{(\ell,h)}$:  
\begin{align}
\{b^{(\ell,h)}\} &= \text{Alloc}(B, L, H, \mathbf{S}) \quad \text{s.t.} \quad \sum_{\ell,h} b^{(\ell,h)} = B.  
\end{align}

\paragraph{Step 3: Selection.} 
Within each head, the top-$b^{(\ell,h)}$ tokens are retained according to their scores:  
\begin{align}
\mathcal{I}^{(\ell,h)} &= \text{TopK}(\mathbf{S}^{(\ell,h)}, b^{(\ell,h)}).  
\end{align}
% \begin{align}
% \mathbf{S} &= \text{Score}(\mathbf{K}, \mathbf{V}, \mathbf{Q}) \in \mathbb{R}^{L \times H \times S} \tag{Scoring}\\
% \{b^{(\ell,h)}\} &= \text{Alloc}(B, L, H) \quad \text{s.t.} \sum_{\ell,h} b^{(\ell,h)} = B \tag{Allocation}\\
% \mathcal{I}^{(\ell,h)} &= \text{TopK}(\mathbf{S}^{(\ell,h)}, b^{(\ell,h)}) \tag{Selection}
% \end{align}
All existing methods follow this top-down ``cake-slicing'' paradigm~\citep{cake}: practitioners first decide a global memory budget $B$, then allocate and select tokens to fit the predetermined size. 
While these methods \emph{can} achieve excellent compression performance under optimal budgets, practitioners face the fundamental challenge of determining \emph{what} is the ``optimal'' budget for each specific inference scenario. 
In practice, it is extremely difficult to directly set the budget to its optimal value, as this requires prior knowledge of task characteristics and attention patterns that are unavailable at deployment time.

\subsection{Existing Works}
\textbf{Scoring Strategies.} A first line of work focuses on designing token-importance \emph{scores} from which the cache is pruned. For example, StreamingLLM~\citep{streamLLM} leverages a sliding-window augmented with attention sinks to approximate query--key affinity. SnapKV~\citep{snapkv} selects important key-value pairs by analyzing the attention from a small tail window of tokens to the prompt, retaining those with high cumulative attention. H2O~\citep{h2o} formulates the selection of important key-value pairs as a heavy hitter detection problem, leveraging algorithms from large-scale data analysis. In practice, these works commonly apply a Top-K rule to select tokens based on their computed scores.

\textbf{Allocation Policies.} A complementary line of research studies how a global memory budget should be \emph{allocated}. The simplest heuristic is a \emph{global Top-$k$} rule that keeps the highest-scoring tokens irrespective of their origin. More refined schemes assign budgets at the \emph{layer} level---PyramidKV~\citep{pyramidkv} uses heuristics to distribute the budget across layers, while SimLayerKV~\citep{simlayerkv} leverages defined layer similarity to guide allocation. At the \emph{head} level, AdaKV~\citep{adakv} achieves theoretical optimality in allocation at the attention score level given a fixed budget, though this does not always translate to optimal end-to-end performance. These works make important contributions to the allocation problem, but they also exacerbate the fundamental challenge: \emph{how should the budget be determined in the first place?}

\section{Method}
\label{sec:method}

GVote departs from the \emph{top--down} ``cake--slicing'' paradigm by determining the \emph{budget itself} rather than optimising within a fixed budget. This section first motivates our bottom--up perspective, then presents the GVote scheme in detail, and finally discusses implementation considerations.

\subsection{Motivation}
\label{subsec:motivation}

What constitutes effective KV-cache compression? We adopt a principled approach based on the \emph{top-$p$} criterion: a compression is considered optimal if it preserves the minimal set of keys whose attention weights contribute to the top-$p$ cumulative probability mass. 
This criterion naturally balances accuracy and memory efficiency by retaining only the most relevant tokens while filtering out negligible contributions.

Consider an idealized scenario where all future queries $\{\mathbf{Q}_t, \mathbf{Q}_{t+1}, \ldots\}$ are known \emph{a priori}. The optimal retention strategy would preserve precisely the \emph{union} of keys required by each subsequent query. 
Formally, the optimal keep-set at step~$t$ is $\bigcup_{u>t} \mathcal{S}_u$, where $\mathcal{S}_u = \arg\text{Top-p}(\mathbf{A}_u)$ denotes the indices selected by the yet-unseen query $\mathbf{Q}_u$ using top-$p$ selection. 
This theoretical perspective clarifies the fundamental objective: the keys preserved during \textbf{prefill} should represent the \emph{aggregate} requirements of \emph{all} future queries, not merely those favoured by the current query.

\begin{figure}[t]
    \centering
    \includegraphics[width=\textwidth]{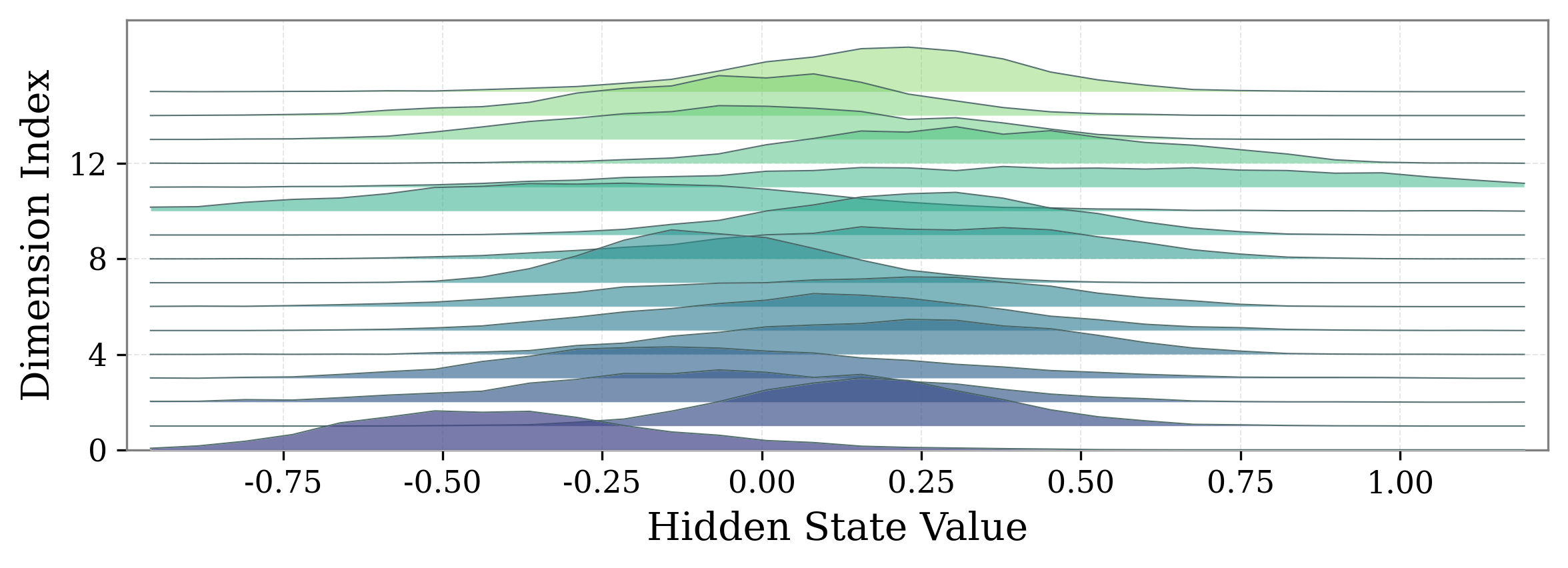}
    \caption{Hidden states exhibit explicit Gaussian structure.}
    \label{fig:gaussian_sampling}
\end{figure}

However, this idealized scenario is unavailable in practice—we cannot know future queries before they are generated. 
How can we estimate these future requirements? A key insight emerges from the statistical properties of transformer hidden states. 
Prior work \citet{expectedattention} discovered that each channel of the hidden state exhibits an approximately Gaussian distribution along the sequence axis, a finding we independently verify in our experiments. This regularity stems from the layer normalization operation, which naturally drives the hidden state distribution toward Gaussianity, as illustrated in Figure~\ref{fig:gaussian_sampling}.

This empirical regularity provides a principled proxy for future query estimation: by sampling from the fitted Gaussian distribution of hidden states, we can synthesize plausible future queries and approximate the union set above. 
GVote operationalizes this intuition through a Monte Carlo approach, where multiple synthetic queries vote on which keys to retain. 
The final keep-set is determined by aggregating votes from all synthetic queries, ensuring robustness across diverse future scenarios.

A subtle but important design choice concerns the selection method for synthetic queries. While our motivation suggests using top-$p$ selection for each synthetic query (consistent with the ``required keys'' definition), we employ top-$k$ instead. This choice reflects the fundamental difference between real and synthetic queries: synthetic queries are merely \emph{estimates} of future queries, leading to noisy attention weights that make top-$p$ selection overly sensitive and prone to including irrelevant tokens. By using top-$k$ with $k = |\mathcal{C}_0|$ (where $\mathcal{C}_0$ is the candidate set from the real current query), we maintain a consistent budget while avoiding the instability of top-$p$ on noisy synthetic attention distributions.

\subsection{The proposed GVote Scheme}
\label{subsec:algorithm}

GVote converts the abstract intuition above into a four--step procedure summarised in Algorithm~\ref{alg:gvote}. All computations are performed \emph{per request} and \emph{per head} and therefore naturally adapt to heterogeneous workloads.

\paragraph{Step\;1: Single–step budget estimation.} At the \textbf{prefill} stage, we first compute the attention weights $\mathbf{A}_0\in\mathbb{R}^{S_\text{head}\times L}$ (Equation~(1)) and apply a nucleus‐style top-$p$ truncation with threshold $p_\text{nuc}$ to obtain a \emph{candidate set} $\mathcal{C}_0$. The size of $\mathcal{C}_0$ provides an estimate of the budget for token selection, guiding each synthetic future query to select a corresponding number of tokens that will later be aggregated.

\paragraph{Step\;2: Hidden–state statistical computation.} We compute the mean $\boldsymbol{\mu}$ and variance $\boldsymbol{\sigma}^2$ of the hidden state $\mathbf{h}_0$ (layer norm output) to parameterise a diagonal Gaussian $\mathcal{N}(\boldsymbol{\mu}, \operatorname{diag}(\boldsymbol{\sigma}^2))$. The first $n_{s}$ tokens are ignored because of the well-known attention sink phenomenon \citep{streamLLM}.

\paragraph{Step\;3: Future query sampling.} We sample a matrix $\widetilde{\mathbf{H}}\in\mathbb{R}^{S\times d_h}$ from the Gaussian and project it into query space via the model's linear layer. We then compute the average of cosine and sine values for $n_f$ future positions and apply rotary positional embedding (RoPE) using this average to obtain $\widetilde{\mathbf{Q}}\in\mathbb{R}^{S\times d_k}$. Each row of $\widetilde{\mathbf{Q}}$ yields its own candidate set using top-$k$ selection with $k = |\mathcal{C}_0|$, which we denote collectively as $\{\mathcal{C}^{(s)}\}_{s=1}^S$.

\paragraph{Step\;4: Voting and aggregation.} The final keep‐set is the union $\mathcal{K} = \bigcup_{s=1}^S \mathcal{C}^{(s)}$, 
% \begin{equation}
%     \mathcal{K} = \bigcup_{s=1}^S \mathcal{C}^{(s)},
% \end{equation}
which determines the effective budget $B = |\mathcal{K}|$. Because $\mathcal{K}$ is derived from multiple plausible future queries, it preserves tokens critical under diverse scenarios while discarding unlikely candidates.

\begin{algorithm}[t]
\caption{GVote: Adaptive Monte--Carlo KV‐cache compression}
\label{alg:gvote}
\small
\begin{algorithmic}[1]
\REQUIRE Keys $\mathbf{K}_{1:L}\in\mathbb{R}^{L\times d_k}$, values $\mathbf{V}_{1:L}$, hidden state $\mathbf{h}_0\in\mathbb{R}^{d_h}$, nucleus threshold $p_\text{nuc}$, number of samples $S$, number of future positions $n_f$
\STATE $\mathbf{A}_0 \gets \operatorname{softmax}\bigl(\mathbf{Q}_0 \mathbf{K}_{1:L}^\top / \sqrt{d_k}\bigr)$ \hfill// $\mathbf{Q}_0$ is current query, $\mathbf{A}_0\!:\! \mathbb{R}^{1\times L}$
\STATE $\mathcal{C}_0 \gets \operatorname{TopP}\bigl(\mathbf{A}_0, p_\text{nuc}\bigr)$ \hfill
\STATE $B_{\text{step}} \gets |\mathcal{C}_0|$ \hfill// step budget
\STATE $\boldsymbol{\mu} \gets \operatorname{mean}(\mathbf{h})$, $\boldsymbol{\sigma}^2 \gets \operatorname{var}(\mathbf{h})$ \hfill// Step 2
\STATE $\widetilde{\mathbf{H}} \sim \mathcal{N}(\boldsymbol{\mu}, \operatorname{diag}(\boldsymbol{\sigma}^2))$ \hfill// $\widetilde{\mathbf{H}}\in\mathbb{R}^{S\times d_h}$
\STATE $\mathbf{P} \gets \operatorname{AverageCosSin}(n_f)$ \hfill// Compute average cos/sin for $n_f$ future positions
\STATE $\widetilde{\mathbf{Q}} \gets \text{RoPE}\bigl(\widetilde{\mathbf{H}} \mathbf{W}_q, \mathbf{P}\bigr)$ \hfill// Apply RoPE using averaged positions, $\widetilde{\mathbf{Q}}\in\mathbb{R}^{S\times d_k}$
\STATE $\mathbf{L} \gets \widetilde{\mathbf{Q}} \mathbf{K}_{1:L}^\top / \sqrt{d_k}$ \hfill// attention logits, $\mathbf{L}\in\mathbb{R}^{S\times L}$
\STATE $\{\mathcal{C}^{(s)}\}_{s=1}^S \gets \operatorname{TopK}\bigl(\mathbf{L}, B_{\text{step}}\bigr)$ \hfill// row‐wise on logits
\STATE $\mathcal{K} \gets \bigcup_{s=1}^S \mathcal{C}^{(s)}$ \hfill// Step 4
\STATE Prune $(\mathbf{K}_{1:L}, \mathbf{V}_{1:L})$ to the indices in $\mathcal{K}$
\RETURN Compressed KV‐cache $(\mathbf{K}_{\mathcal{K}}, \mathbf{V}_{\mathcal{K}})$
\end{algorithmic}
\end{algorithm}

\subsection{Are synthesized queries good approximations?}
\label{subsec:approximation_quality}
The efficacy of GVote hinges on a central hypothesis: that synthetic queries, sampled from the statistical distribution of hidden states, can serve as a reliable proxy for the model's \emph{actual} future queries. But how valid is this approximation? In this section, we present empirical evidence to validate this core assumption.

\begin{figure}[t]
    \centering
    \subfloat[Attention overlap and token usage analysis]{
        \includegraphics[width=0.85\textwidth]{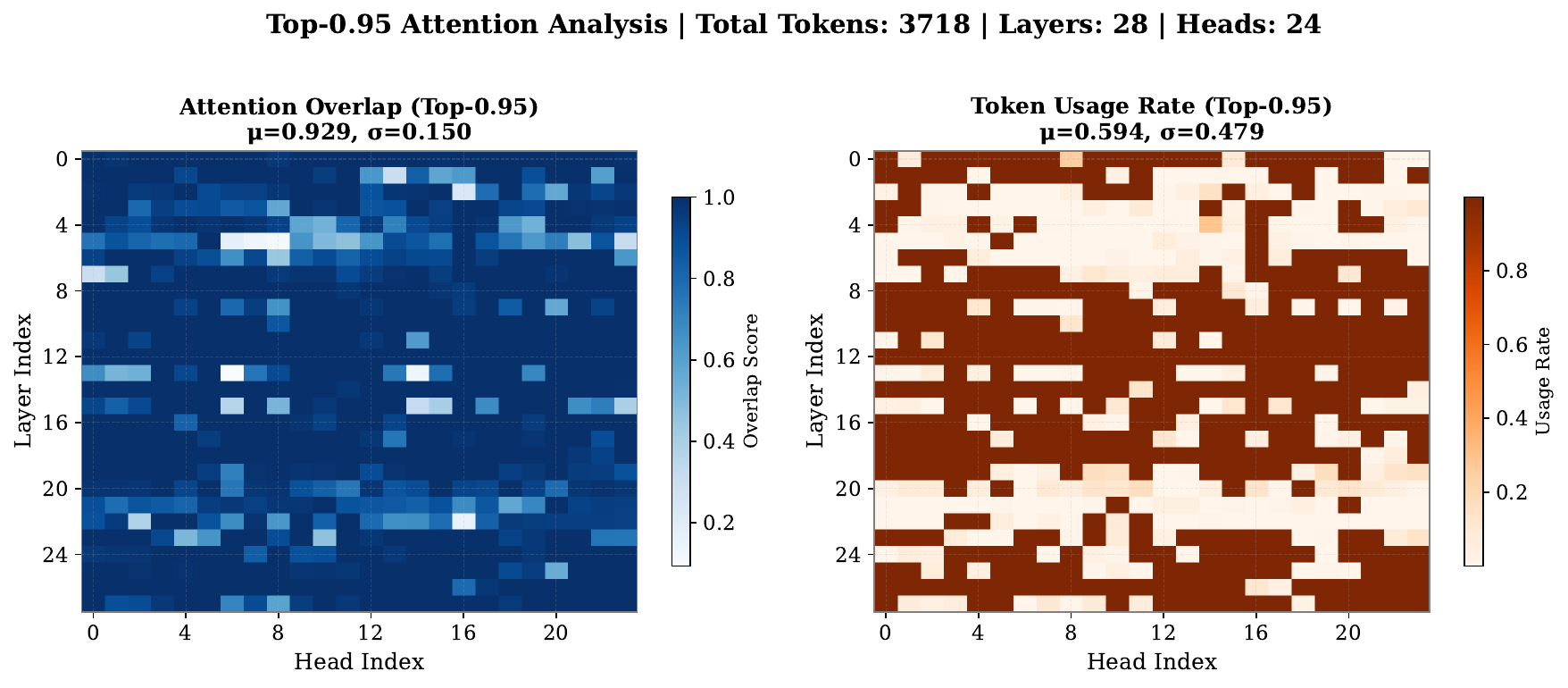}
        \label{fig:attention_analysis}
    }
    \vfill
    \subfloat[Attention correlation analysis]{
        \includegraphics[width=0.85\textwidth]{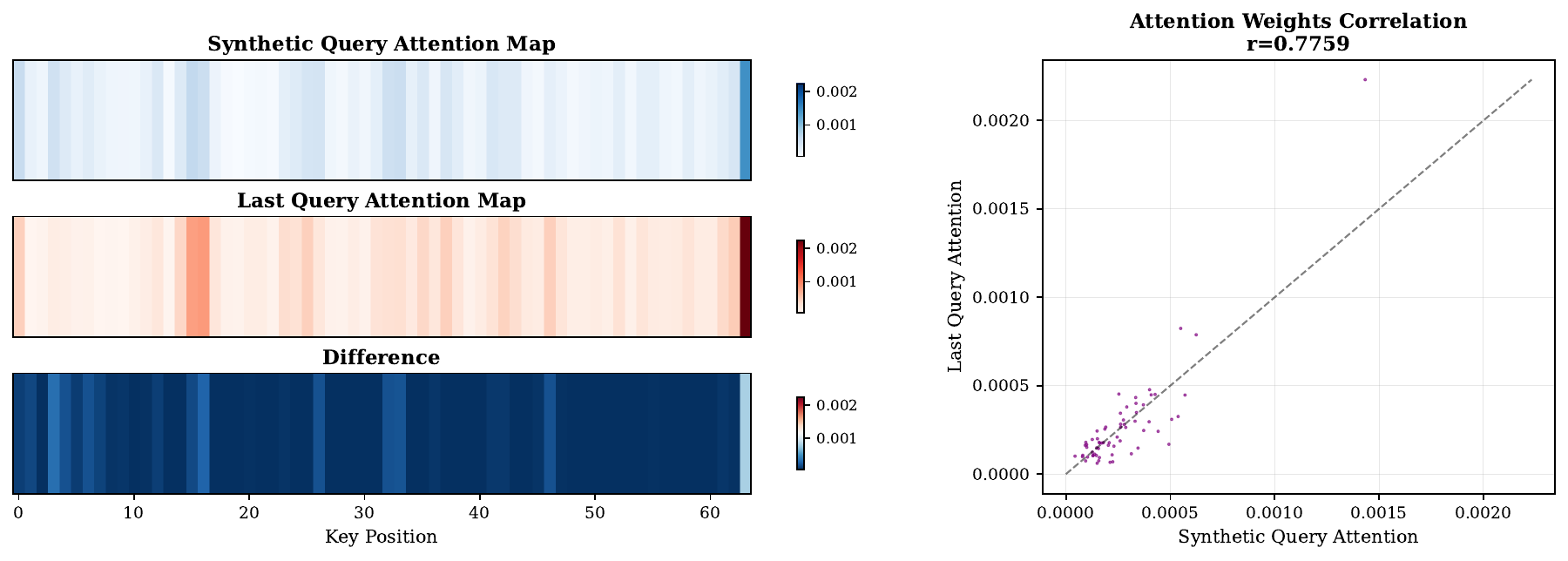}
        \label{fig:correlation_analysis}
    }
    \caption{Comprehensive analysis of synthetic query approximation quality. \subref{fig:attention_analysis} Layer-wise and head-wise attention overlap and token usage rates, showing consistency between synthetic and real query attention patterns with high overlap scores (mean $\mu=0.929$) and moderate token usage rates (mean $\mu=0.594$). \subref{fig:correlation_analysis} Detailed attention correlation analysis for a specific attention head, including attention map comparison and correlation scatter plot with $r=0.7759$, demonstrating strong alignment between synthetic and real query attention patterns.}
    \label{fig:attn_map_comparison}
\end{figure}

To investigate this, we first compare the attention patterns generated by synthetic queries against those from real, subsequent queries. Our experimental setup works as follows: given a sentence, we mask the last token and generate a single synthesized query using our method. We then compute attention weights for both the synthesized query and the ground truth query (the last real token) against the entire sentence. Figure~\ref{fig:attn_map_comparison} provides a comprehensive analysis of attention correlation between synthetic and real queries. The left panel shows layer-wise and head-wise attention overlap and token usage rates, while the right panel provides a detailed example of attention map correlation for a specific attention head.

We define \emph{Attention Overlap} as follows: for each synthesized query, we select the top-0.95 attention weights to form a candidate set. The overlap score is then computed as the sum of attention weights in the ground truth query that correspond to the same tokens in this candidate set. This metric quantifies how well our synthesized queries capture the salient tokens that the model actually attends to in the future.

We observe a strong qualitative correspondence between the aggregated attention scores from synthetic queries and the attention scores from the ground-truth future query, with a strong Pearson correlation coefficient of $r=0.7759$. Both tend to focus on similar token regions, suggesting that our sampling process captures the salient features that are likely to be attended to in the future.
However, closer examination reveals limitations in single-query synthesis. The attention overlap of $0.929$ falls slightly below our $0.95$ threshold, indicating inclusion of some less critical tokens. Moreover, we also investigated the actual token requirement of ground truth query and found it needs only $\sim40\%$ of tokens, highlighting potential redundancy.

These issues arise from variance in synthetic queries. Multiple sampling mitigates variance, but at the cost of including more irrelevant tokens. 
Our approach is guided by a key insight: the size of the union of tokens required by a set of future queries is necessarily greater than or equal to the size of tokens required by any single query within that set.
GVote operationalizes this by using the current query's budget ($|\mathcal{C}_0|$) as a proxy for a single future query's need. Each synthetic query is then constrained to select this many tokens via top-$k$. 
This strategy, combined with the final union, ensures that the keep-set is robust to the noise of individual samples while controlling for redundancy.

\subsection{Implementation and Overhead}
\label{subsec:implementation}

The GVote algorithm is designed for efficient, parallel execution on GPU architectures. In our PyTorch-based implementation, the core logic (Algorithm~\ref{alg:gvote}, lines 5--10)---spanning synthetic query generation, RoPE application, and attention score computation---is fully vectorised. The $S$ synthetic queries are processed in a single batch, amortising the computational cost. The union operation (line 11) is performed efficiently using boolean masks; instead of materialising explicit index lists, a mask is computed for each of the $S$ candidate sets. These masks are aggregated with a single logical OR operation (\texttt{torch.any}), and the final indices for $\mathcal{K}$ are extracted via a \texttt{nonzero} operation. The subsequent pruning (line 12) is then a simple indexing operation which gathers the selected keys and values. This entire procedure introduces a one-time computational overhead during the prefill phase. While non-trivial, this cost is fixed and does not scale with the number of generated tokens.

During the decoding phase, the GVote procedure is applied independently for each attention head, leading to a \emph{non-uniform} KV-cache where different heads may retain a varying number of key-value pairs. This structurally sparse format precludes storage as a single dense tensor. However, it is fully compatible with modern attention kernels like FlashAttention \citep{flashattention}, which are designed to handle variable-length sequences via the \texttt{varlen} interface. This approach of creating a non-uniform, content-aware cache is consistent with prior work like AdaKV \citep{adakv}.

\subsection{Parameters}
\label{subsec:parameters}

The GVote algorithm introduces two key hyperparameters: the nucleus sampling threshold $p_\text{nuc}$ and the number of synthetic queries $S$.

\paragraph{Nucleus threshold ($p_\text{nuc}$).} This parameter (Algorithm~\ref{alg:gvote}, line~2) sets the cumulative probability mass of attention weights that must be preserved—effectively specifying the recall of the softmax attention distribution. We recommend $p_\text{nuc}=0.95$ in practice, which offers decent accuracy-memory trade-off. For short sequences $p_\text{nuc}=0.95$ would suggests a slightly larger proportion, yet the effect is trivial since the absolute length of the sequence is small. Detailed sensitivity analysis is presented in [ref].

\paragraph{Number of samples ($S$).} This parameter (Algorithm~\ref{alg:gvote}, line 5) controls the number of synthetic future queries used to form the final keep-set $\mathcal{K}$. In practice, the impact of $S$ on the final budget is less pronounced than that of $p_\text{nuc}$. While a larger $S$ can produce a more robust estimate, it typically comes with higher KVCache usage and higher computational overhead. We find that $S \ge 8$ typically offers a good compromise between robustness and efficiency. We recommend using a conservative $p_\text{nuc}$ with a moderately high $S$. Further empirical results are discussed in the experiments section~[ref].
\section{Experiments}
\label{sec:experiments}

We conduct comprehensive experiments to evaluate GVote's effectiveness across diverse tasks and compare it with state-of-the-art KV-cache compression methods. Our experiments demonstrate that GVote achieves superior accuracy-memory trade-offs without requiring manual budget tuning.

\subsection{Experimental Setup}
\label{sec:exp_setup}

\textbf{Datasets.} We evaluate on eight diverse benchmarks spanning different sequence lengths and task types: GSM8K (mathematical reasoning)\cite{gsm8k}, RULER-CWE \cite{ruler}, and Longbench\cite{longbench2} (long-context understanding).

\textbf{Baselines.} We compare against three representative KV-cache compression methods:
\begin{itemize}
    \item \textbf{StreamLLM}~\cite{streamLLM}: Streaming attention with sink tokens and fixed window.
    \item \textbf{SnapKV}~\cite{snapkv}: SnapKV utilizes the last tokens to score the KVCache of previous context.  
    \item \textbf{AdaKV}~\cite{adakv}: Adaptive  KV-cache compression across different heads. It proves its optimal allocation strategy given fixed budget, yet the budget itself is pre-determined
\end{itemize}

\textbf{Model and Implementation.} We extensively tests four popular models accross different families and sizes: Llama3.1-8B-Instruct, Llama3.2-3B-Instruct\cite{llama3}, Qwen2.5-7B-Instruct, Qwen2.5-14B-Instruct \cite{qwen2}.

\textbf{Evaluation Metrics.} We measure accuracy and effective cache usage ratio. For each baseline, we sweep compression ratios from 10\% to 50\% to construct accuracy-usage curves. GVote operates without manual budget specification, automatically determining the optimal compression ratio per request.

\subsection{Main Results}
\label{sec:main_results}

\begin{figure}[t]
    \centering
    \includegraphics[width=\textwidth]{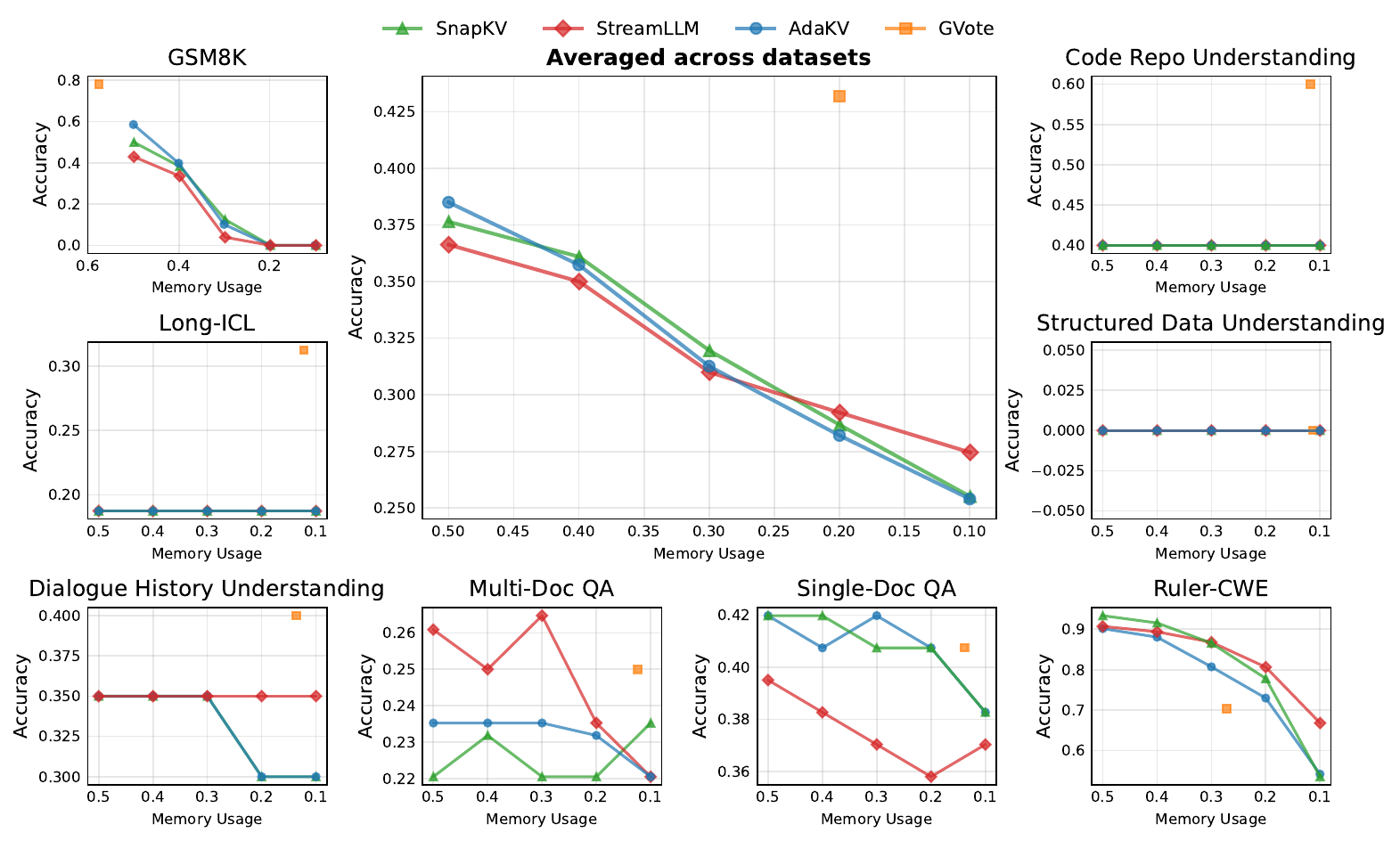}
    \caption{Accuracy vs. Cache Usage across eight benchmarks using Qwen2.5-7B-Instruct. Each baseline shows results across different compression ratios (10\%-50\%). The optimal budgets across different datasets are various, a characteristics that fixed budget cannot deal with. GVote (orange square) consistently achieves accuracy-usage trade-off sweet spot compared to baselines with fixed budgets. }
    \label{fig:main_results}
\end{figure}

Figure~\ref{fig:main_results} presents the core results across all eight datasets and reports the average performance. GVote achieves higher accuracy under significantly low budget under most scenarios, and hence the overall performance is impressive, reducing 2x memory budget while maintaining the accuracy.

\subsection{Multi-Model Evaluation}
\label{sec:multi_model}

\begin{figure}[t]
    \centering
    \includegraphics[width=\textwidth]{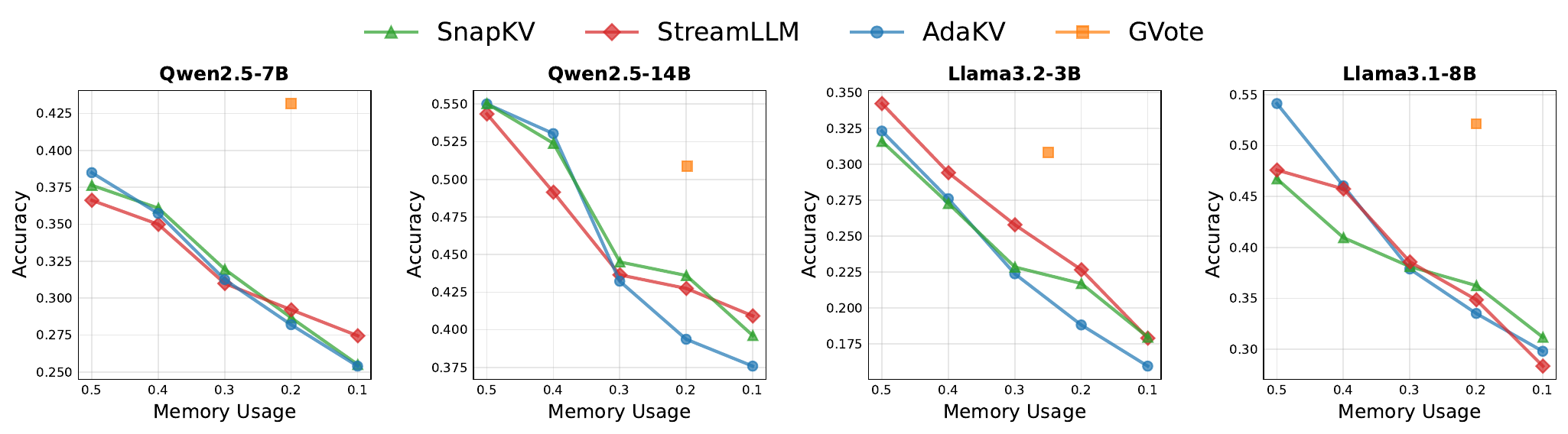}
    \caption{GVote performance across different model architectures and sizes. All models are of Instruct version. We omit the suffix due to space limitation. The figure shows accuracy-usage curves for various models, demonstrating the adaptive mechanism's generalization across different model configurations.}
    \label{fig:multi_model}
\end{figure}

Figure~\ref{fig:multi_model} demonstrates GVote's effectiveness across different model architectures and sizes. The adaptive compression mechanism maintains its advantages across various model configurations, indicating good generalization properties.

\subsection{Hyperparameter Analysis}
\label{sec:hyperparameter}

In this section, we investigate about what is the optimal hyperparameters of GVote. We primarily test different parameter combinations on RULER dataset.

\begin{figure*}[t]
    \centering
    \begin{minipage}[t]{0.45\textwidth}
        \centering
        \includegraphics[width=\textwidth]{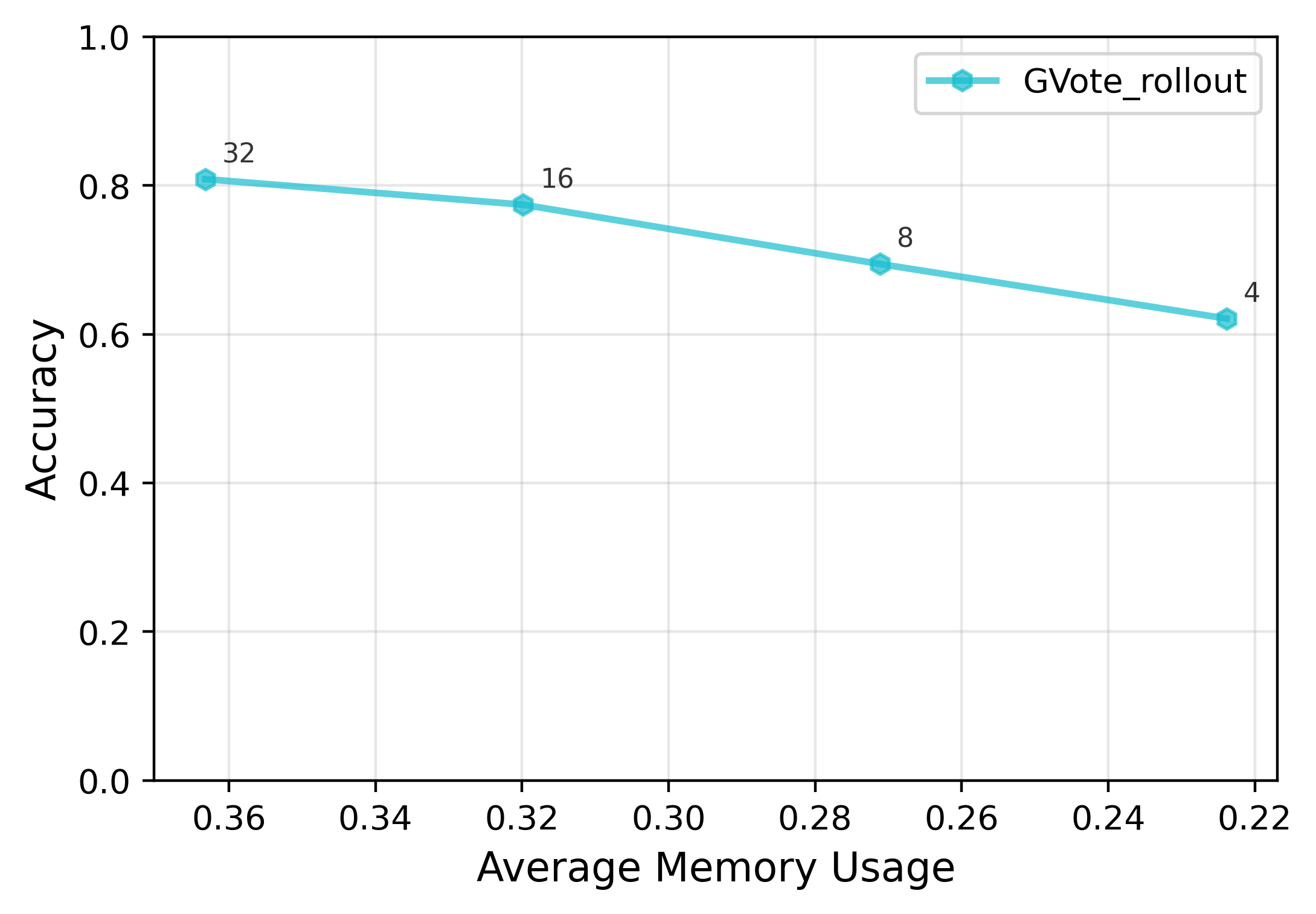}
        \caption{Effect of sampling number $S$ on GVote performance. Higher $S$ would deliver better accuracy with moderately higher usage. However it would cost more memory and computation for compression.}
        \label{fig:sampling_analysis}
    \end{minipage}
    \hfill
    \begin{minipage}[t]{0.45\textwidth}
        \centering
        \includegraphics[width=\textwidth]{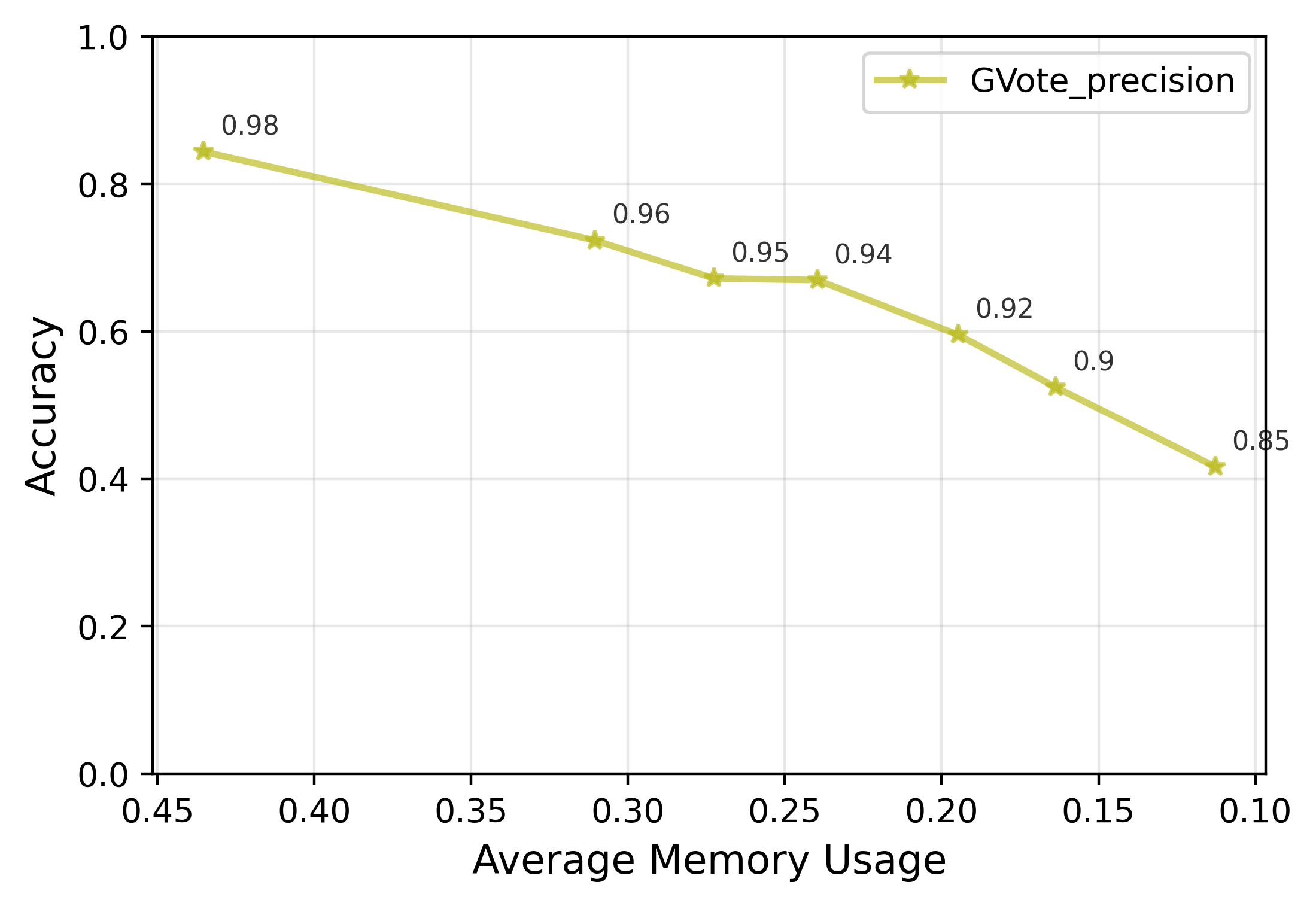}
        \caption{Effect of nucleus sampling threshold $p_{nuc}$. Similarly, higher $p_{nuc}$ would result in a higher accuracy and memory usage. However, it seems to be more sensitive compared to $S$.}
        \label{fig:nucleus_p_analysis}
    \end{minipage}
    \label{fig:combined_analysis}
\end{figure*}

\textbf{Sampling number $S$.}
Figure~\ref{fig:sampling_analysis} demonstrates the effects of different sampling number $S$ on final performance. As intuition, higher $S$ would approximate the real distribution better. However, in case the context length is huge, for example, $>128K$, higher $S$ would result in an extremely large intermediate logits matrix $\mathbf{L}$, potentially leading to out of memory errors.

\textbf{Nucleus Sampling Threshold ($p_{nuc}$).} Figure~\ref{fig:nucleus_p_analysis} examines how the nucleus sampling threshold for single-step budget estimation affects overall performance. Similar to $S$, higher $p_{nuc}$ would also result in both high accuracy and memory usage. However, it is more sensitive to $S$, as it reaching comparable accuracy with higher memory. The specific value of $p_{nuc}$ has no impact on pruning speed, unlike $S$.

\section{Conclusion}

We observe the uneven budget requirements of different requests and hence propose GVote to address this challenge. By moving beyond a one-size-fits-all approach and embracing the heterogeneous sparsity of requests through a Gaussian distribution, we have shown a clear path to substantial memory reduction. Our findings open the door for exciting future work. Specifically, we aim to investigate even more sophisticated adaptive methods that eliminate the need for an accuracy-memory trade-off. We also look forward to tackling the critical engineering challenges required to bring these theoretical gains into practical application.

\section{LLM Usage}

The ideation, testing, and research process for this paper were human-led, with assistance from a Large Language Model (LLM) for refinement. Specifically, the manuscript was initially drafted by humans, after which an LLM was used for grammar and formatting checks. Furthermore, the experimental data was collected manually by humans, while its visualization was created with the assistance of the LLM.

\bibliography{ref}
\bibliographystyle{iclr2026_conference}

\end{document}